# Geo-located Twitter as the proxy for global mobility patterns

Bartosz Hawelka[a,b,*], Izabela Sitko[a,b], Euro Beinat[a], Stanislav Sobolevsky[b], Pavlos Kazakopoulos[a] and Carlo Ratti[b]

[a] *GISscience Doctoral College, Department of Geoinformatics – Z_GIS, University of Salzburg, Austria*
[b] *SENSEable City Laboratory, Massachusetts Institute of Technology, Cambridge, USA*
*\* Corresponding author: bartosz.hawelka@sbg.ac.at*

In the advent of a pervasive presence of location sharing services researchers gained an unprecedented access to the direct records of human activity in space and time. This paper analyses geo-located Twitter messages in order to uncover global patterns of human mobility. Based on a dataset of almost a billion tweets recorded in 2012 we estimate volumes of international travelers in respect to their country of residence. We examine mobility profiles of different nations looking at the characteristics such as mobility rate, radius of gyration, diversity of destinations and a balance of the inflows and outflows. The temporal patterns disclose the universal seasons of increased international mobility and the particular national nature of overseas travels. Our analysis of the community structure of the Twitter mobility network, obtained with the iterative network partitioning, reveals spatially cohesive regions that follow the regional division of the world. Finally, we validate our result with the global tourism statistics and mobility models provided by other authors, and argue that Twitter is a viable source to understand and quantify global mobility patterns.

## Introduction

Reliable and effective monitoring of the worldwide mobility patterns plays an important role in studies exploring migration flows (Castles and Mill 1998; Greenwood 1985; Sassen 1999), touristic activity (Miguéns and Mendes 2008) but also the spread of diseases and epidemic modeling (Bajardi et al. 2011; Balcan et al. 2009). Traditionally, those studies relied either on the aggregated and temporally-sparse official statistics or on the selective small-scale observations and surveys. In a more recent approach, international mobility was approximated with the air traffic volumes (Barrat et al. 2004), i.e. the dataset with the potentially global coverage but biased toward just one mode of transportation and in many cases difficult to obtain. However, in the advent of a pervasive presence of location sharing services researchers have gained an unprecedented access to the direct records of human activity. Each day, millions of individuals leave behind their digital traces by using services such as mobile phones, credit cards or social media. Most of those traces can be located in space and time, and thus constitute a valuable source for human mobility studies.

Out of several types of collectively sensed data, the one that was the most intensively explored for analysis of human mobility has undoubtedly been cellular phone records. With truly pervasive character they provided a basis to formulate important findings on the nature of a collective movement in urban (Calabrese et al. 2010; Kang et al. 2012), regional (Sagl et al. 2012; Calabrese et al. 2013) and country-wide scales (Krings et al. 2009; Simini et al. 2012). Nevertheless, a high fragmentation of the mobile telecom market actually excludes the availability of any worldwide dataset. In this case, a good alternative is offered by social media data. Despite its lower penetration and a potential bias towards a younger part of the population, social media constantly gains in popularity and representativeness (Gesenhues 2013) and in most cases are global by design.

In this study we attempt to uncover global mobility patterns, as well as to compare mobility characteristics of different nations. Our work is based on the data from Twitter – one of the most popular social media platforms, with over 500 million users registered by mid-2013 (Twitter Statistics 2013). Initially established in the USA, the service has quickly spread to other countries (Java et al. 2007; Leetaru et al. 2013; Mocanu et al. 2013), becoming a worldwide phenomenon. By design Twitter is an open and public medium, which practically limits privacy consideration, especially in studies such as ours, which seek for collective rather than individual patterns of human behavior. In particular, we take advantage of the portion of tweets augmented with explicit geographic coordinates, as measured either by GPS embedded in a mobile device, or located to the nearest address based on the IP location of a computer. We call these tweets geo-located tweets. They are still a limited sample of all tweets and account for around 1% of the total feed (Morstatter et al. 2013). However, thanks to the increasing penetration of smart devices and mobile applications, the volume of the geo-located Twitter is constantly growing (Figure 1) and progressively becoming a valuable register of human traces in space and time. The absolute volume of 3.5M geo-located tweets per day (authors' calculation for December 2012) appears as a promising base for carrying out the worldwide mobility analysis, which as such is also the subject of our exploration.

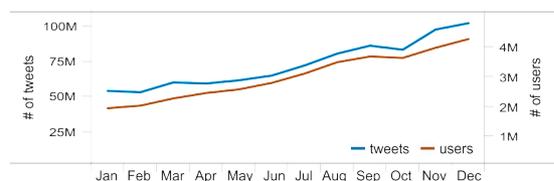

Figure 1. Number of geo-located tweets (blue line) and users (orange line) per month in 2012.

Because of the richness yet simplicity of the medium, Twitter has already been the subject of many studies for a variety of applications. First explorations focused on the properties of Twitter as a social network, already proving its global character and scientific potential one year after the launch of the service (Java et al. 2007; Kwak et al. 2010; Huberman et al. 2008). Another line of research examined a content of tweets to assess a mood of society (Golder and Macy 2011; Bollen et al. 2011; Pak and Paroubek 2010), recently also with the geographic perspective based on a geo-located stream (Frank et al. 2013, Mitchell et al. 2013). The next area that has received much attention was crisis management (MacEachren et al. 2011; Thom et al. 2012; Sakaki et al. 2010), where the emphasis was placed on the detection of anomalous activity, as well as on the potential of a locally generated content to inform emergency services. Geo-located Twitter data were also considered as a support in urban management and planning (Wakamiya et al. 2011; Frias-Martinez et al. 2012), as well as in public health assessment (Ghosh and Guha 2013). All the aforementioned works were spatially selective, focusing on specific study areas. The global perspective was introduced in the study of Kamath et al. (2012), with the analysis of the geographic spread of hashtags. Furthermore, Leetaru et al. (2013) attempted to describe the geography of Twitter based on a one-month sample of global geo-located tweets, while Mocanu et al. (2013) described the global distribution of different languages used while tweeting.

Given the well acknowledged role of location information registered within social networking services, the attempts to translate it into mobility characteristics remain relatively sparse. The most important foundations were provided by Cheng et al. (2011), who analyzed different aspects of mobility based on Twitter check-ins, at that point dominated by the feed from another location sharing service i.e. Foursquare. The study had an extensive scope, but was limited by the time data availability. Other explorations were given e.g. by Cho et al. (2011), who modeled the influence of human mobility on social ties in Gowalla and Brightkite services, or by Noulas et al. (2012), who focused on intra-urban mobility approximated with Foursquare check-ins.

In this paper, we present a global study of mobility. We focus on the worldwide patterns that emerge from the analysis of Twitter data, as well as on the mobility characteristics of different nations. Furthermore, we seek to discover spatial patterns and clusters of regional mobility. Finally, we attempt to validate the representativeness of geo-located Twitter as a global source for mobility data. The paper is organized as follows. First of all, we describe the dataset and illustrate a method to assign users to a country of residence, hence allowing the determination of home users and foreign visitors. Next, we present and compare mobility profiles of various countries, as well as the temporal patterns of inflows and outflows dynamics. Further, we explore a country-to-country network of the flows of travelers, and perform a community detection delineating global regions of mobility. Finally, we validate the results in two ways: (i) comparing against the worldwide tourism statistics and (ii) comparing Twitter data against commonly used models of human mobility.

**Data preparation and pre-processing**

Our study relies on one full year of geo-located tweets, which were posted by users all over the world from January $1^{st}$ until December $31^{st}$ 2012. By geo-located we mean the messages with explicit geographic coordinates attached to each message. The database consists of 944M records generated by a total of 13M users. The stream was gathered through the Twitter Streaming API[1]. Although the service sets a limit on the accessible volume to less than 1% of the total Twitter stream, the total geo-located content was found not to exceed this restriction (Morstatter et al. 2013). Therefore we believe that we successfully collected the complete picture of global geo-located activity within 2012.

Before the actual analysis, the database had to be cleaned from evident errors and artificial tweeting noise, which could pollute mobility statistics. First we examined all the consecutive locations of a single user, and excluded those that implied a user relocating with a speed over 1000km/h, i.e. faster than a passenger plane. Further, we filtered out non-human Twitter activity such as web advertising (e.g. tweetmyjob), web gaming (e.g. map-game) or web reporting (e.g. sandaysoft). Those services can generate significant volumes of data, which does not reflect any sort of human physical presence in the reported place and time. To correct for this noise we checked the popularity of a message's source, assuming that those with only few users can probably be classified as an artificial activity. As the threshold we used a cumulative popularity among 95% of users, constructing the ranking separately for each country. All tweeting sources falling below the threshold were discarded from further analysis. In total, the refinement procedure preserved 98% of users and 95% of tweets from the initial database.

---

[1] https://dev.twitter.com/docs/streaming-apis

**Definition of a country of the user's residence**

An essential first step in our cross-country mobility analysis was an explicit assignment of each user to a country of residence. This made our work different from most of the other Twitter studies, which usually did not attempt to uncover users' origin and characterized a study area using only the total volume of tweets observed in this area (e.g. Mocanu et al. 2013). While for certain research problems this approach is suitable, from the perspective of a global mobility study the differentiation between residents and visitors is crucial. It enables a clear definition of origin and destination of travels and reveals which nation is traveling where and when. Taking advantage of the history of tweeting records of every user, we defined her country of residence as the country where the user has issued most of the tweets. Once the country of residence was identified, the user's activity in any other region of the world was considered as traveling behavior, and the user was counted as a visitor to that country.

We use the country definitions of the Global Administrative Areas spatial database (Global Administrative Areas 2012), which divides the world into 253 territories. Twitter "residents" were identified in 243 of them, with the number of users greatly varying among different countries. The unquestionable leader is USA with over 3.8M users, followed by the United Kingdom, Indonesia, Brazil, Japan and Spain with over 500K users each. There are also countries and territories with only few or no Twitter users assigned.

To evaluate the representativeness of Twitter in a given area, a more illustrative metric is the penetration rate, defined as the ratio between the number of Twitter users and the population of a country. As expected, this ratio does not distribute uniformly across the globe and scales superlinearly with the level of a country's economic development approximated by a GDP per capita (Figure 2A and B). While this property has already been described e.g. by Mocanu et al. (2013) the goodness of a fit of a power law approximation increased when considering penetration of only residents rather than all Twitter users appearing in a country. In the analysis we exclude all countries with a penetration rate below 0.05% (we also exclude countries with the number of resident users smaller than 10,000).

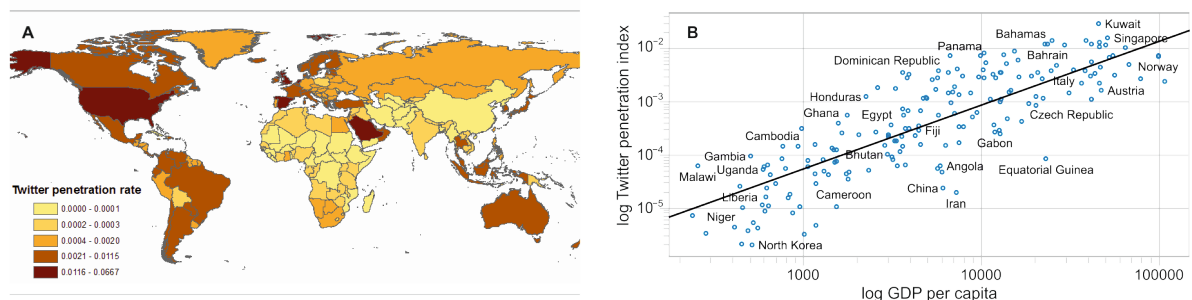

Figure 2. Twitter penetration rate across countries of the world. (A) Spatial distribution of the index. (B) Superlinear scaling of the penetration rate with per capita GDP of a country. $R^2$ coefficient equals 0.70.

**Mobility profiles of countries**

Human mobility can be analyzed at different levels of granularity. In this study we considered a user as being 'mobile' if over the whole year the user had been tweeting from at least one country other than her country of residence. In total, this applied to 1M users, around 8% of all those who used geo-located Twitter in 2012. Figure 3 shows the percentage of mobile users per country and the (geo-located) Twitter penetration in that country in 2012. Most of the top mobile countries, e.g. Belgium, Austria, were characterized by only moderate levels of Twitter adoption. On the other hand users of geo-located Twitter from the USA, the country with the highest penetration rate, revealed a surprisingly small tendency to travel. The only two countries with a high mobility and penetration rates were Singapore and Kuwait. In general, while an increased popularity of Twitter can be treated as a sign of a more active society, it did not immediately imply higher mobility of its users.

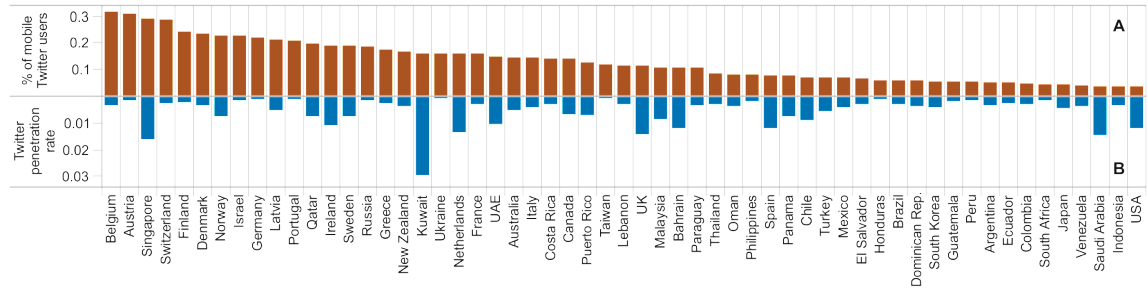

Figure 3. Top countries with the highest rates of users' travel activity.

Next, we examined how spatially spread or concentrated the mobility of users is in a certain nation. This was captured through the average radius of gyration of the users. The radius of gyration measures the spread of user's locations around her usual location. Here, we defined a usual location as the center of mass rather than a home location, as the latter is defined broadly - only with the assignment to a particular country. For each user, the radius of gyration was computed according to the equation:

$$r_g = \sqrt{\frac{1}{n}\sum_{i=1}^{n} |\bar{a}_i - \bar{a}_{cm}|^2} \qquad (1)$$

where *n* is the number of tweeting locations, $\bar{a}_i$ represents the location of a particular tweet (a pair of xy coordinates), and $\bar{a}_{cm}$ is a user's center of mass. Low values of the radius indicate a tendency to travel locally, while bigger values indicate more long-distance travels. The average values computed for users from different countries are shown in Figure 4. A first observation indicated an obviously important role played by the geographical location of a country. Isolated countries such as New Zealand or Australia had an average radius of gyration of over 700km. There was also a positive correlation between the average distance travelled by residents of a country and the mobility rate of its Twitter population, as well as the number of visited countries (Figures 4A and 4B). In any case, the drivers for increased mobility are connected with the economic prosperity of a country, as all received rankings were led by highly developed countries.

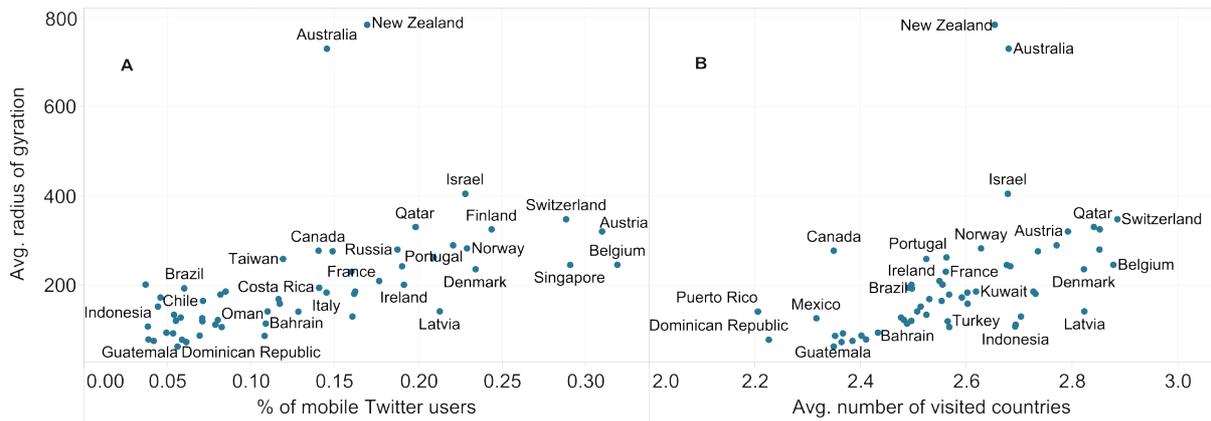

Figure 4. Average radius of gyration of users from different countries compared to (A) percentage of mobile Twitter users and (B) number of countries visited.

The mobility profile of each country can be analyzed from two perspectives: a country being the origin or the destination of international travels. By building the directional country-to-country network of the human flows, we were able to quantify both the inflow and outflow of visitors. Figure 5 shows the analysis for each country based on Twitter users and on total number of Twitter travelers normalized by the Twitter penetration rate in this country, used as an estimation of a total mobility flux. Figures 5A and 5B show the number of Twitter users originating from a country and traveling to another (5A) and users visiting this particular country (5B). Figures 5C and 5D present number of Twitter travelers normalized by the Twitter penetration rate in the country of a user's origin. In case of international arrivals, both the raw number of Twitter users and the estimated population of visitors unveiled the most visited countries to be USA, UK, Spain and France. On the other hand, the ranking of travelers' nationality seemed highly influenced by Twitter's penetration rate, with the biggest groups coming from the countries of high Twitter popularity. Furthermore, low penetration indices leaded to an overestimation of the actual travelers' volume, which may explain the values estimated for Russia or Germany. Figure 5E presents the yearly balance between the estimated inflow and outflow of travelers revealing different countries to be either the origin or destination of international trips.

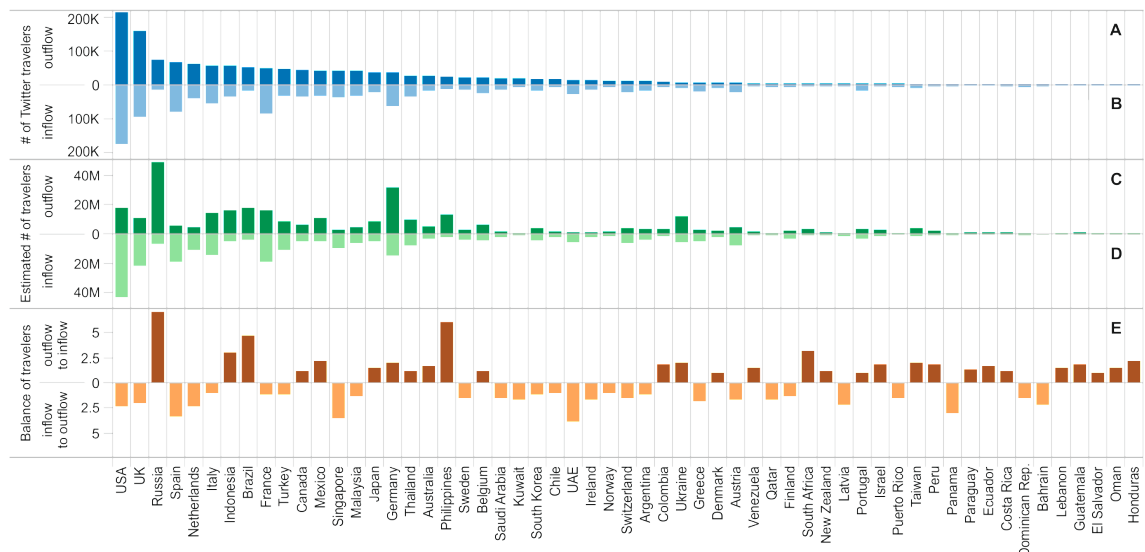

Figure 5. Number of visitors coming from or arriving in a country. Number of Twitter travelers (A and B), number of Twitter travelers normalized by the Twitter penetration rate in the country of origin of the visitor (C and D) and the overall balance of travelers (E).

**Temporal patterns of mobility**

Human mobility is always subject to temporal variations. In order to uncover patterns occurring at the global, as well as at the country level, we measured how many Twitter users were active outside of their country of residence for each day of 2012. The first pattern that emerged from the analysis was the weekly scheme of check-ins abroad (Figure 6). The tendency of increased mobility over weekends seemed to be universal across the globe. Moreover, there were two obvious seasons of higher mobility: the summer months of July and August, and the end of the year, connected to Christmas and New Year's Eve holidays.

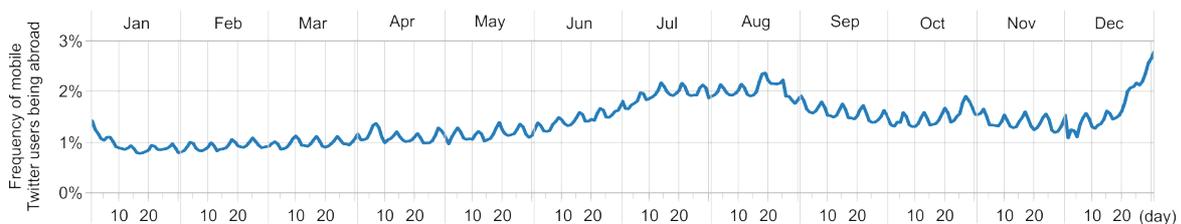

Figure 6. Global temporal pattern of users who traveled abroad.

Looking at the specific countries we discovered a variety of deviations from the aforementioned global pattern. Several of them were easy to interpret and shared by more than one country. For instance, there was a substantial group of European nations with the biggest peak over one of the summer months and a few smaller ones, most probably connected to extended weekends e.g. at the beginning of May (examples are shown in Figure 7A). Another group exhibited a similar pattern, however with the summer mobility increase stretched between June and September (Figure 7B). An interesting example of the mobility behavior influenced by the social and cultural features of a country was observed in the group of Arabic countries (Figure 7C). The period of Ramadan corresponded to a major decrease in the amount of travels abroad, while the time of the Mecca pilgrimage marked a sharp peak at the end of October. In all cases, the end of the year corresponded to a time of increased international mobility.

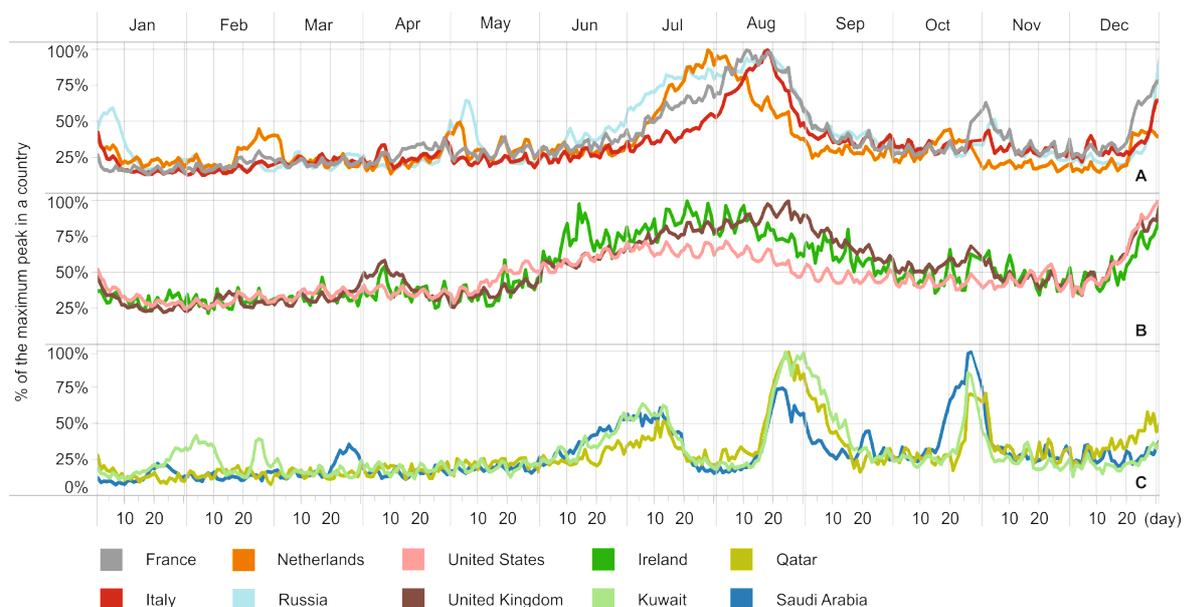

Figure 7. Normalized temporal patterns of mobility, from the perspective of the country of origin. The values for each country are scaled between 0 and 100% of the maximum daily number of travelers being abroad during 2012.

The temporal variations of the inflow of visitors were much more stable than the outflow patterns. Visual inspection of those patterns indicated three main groups. The first group included countries without any specific seasonality of international visits (with the exception of the end of the year period). The second group covered popular summer touristic destinations such as Spain, Italy, Croatia or Greece (Figure 8), with a significant increase in arrivals over the months of July and August. Finally, in the third group we included countries where increased international visits were connected to special events such as Euro 2012 in Poland or the 2012 Olympics in United Kingdom.

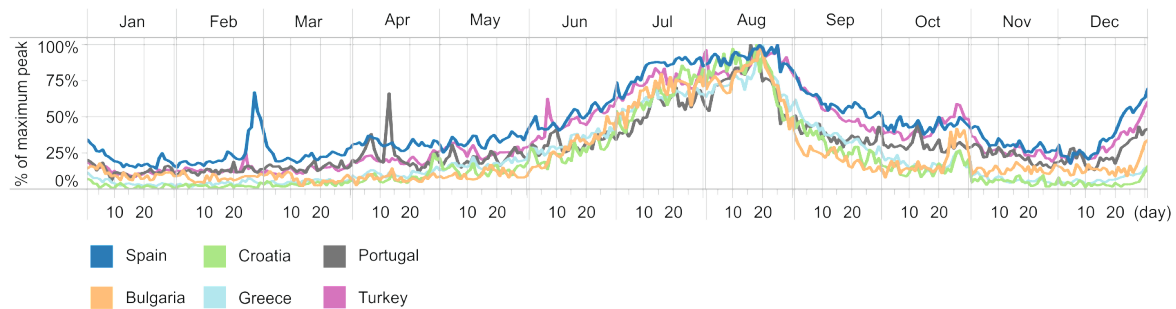

Figure 8. Examples of summer destinations of touristic activity marked by the increased inflow of international Twitter users over summer. The values for each country are scaled between 0 and 100% of the maximum daily number of international visitors during 2012.

**Country-to-country network & partitioning**

In the next step, we analyzed the topology of the country-to-country mobility network created by travelers estimated through the Twitter community. As it has already been proven by many studies, partitioning of a raw network of human communicational interactions e.g. based on mobile phone data (Ratti et al. 2010; Blondel et al. 2010; Blondel 2011; Sobolevsky et al. 2013a), as well as partitioning of human mobility (Amini et al. 2013, Kang et al. 2013), can lead to the delineation of spatially cohesive communities, aligning surprisingly well with the existing socio-economic borders of the underlying geographies. Our aim was to test if this finding holds true for the Twitter-based mobility network, and if so, which distinctive mobility clusters emerge in different parts of the world.

Taking advantage of our methodology of assigning a user to their country of residence and focusing on the mobile users, we built the worldwide country-to-country network of Twitter flows. Each country was considered as a node of the network, and the edges were weighted with the number of Twitter users travelling between a pair of nodes. The network was directional, as the connections were built from the country of residence to each other country where a user appeared as a visitor. To deal with the sparseness of the network and different levels of Twitter representativeness, we filtered out all countries with the outgoing population smaller than 500 Twitter users, as well as those countries where the Twitter penetration was below 0.05%. Furthermore, the flows were normalized by the Twitter penetration rate in the country of a user' origin in order to estimate the real mobility flux rather than just a number of Twitter users. The top 30 flows between different countries are presented in Figure 9.

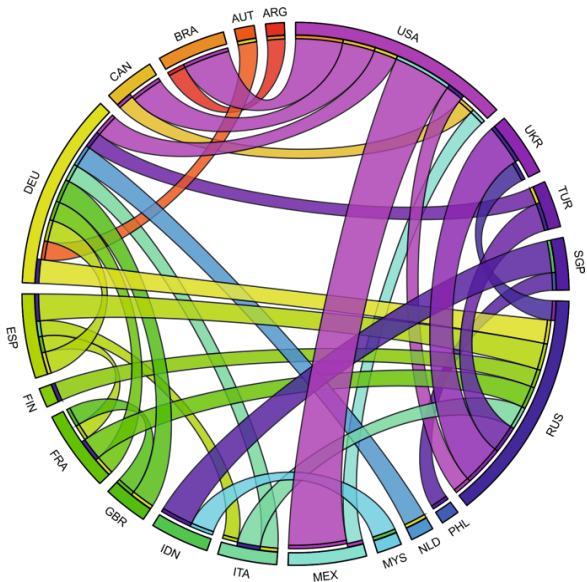

Figure 9. Top 30 country-to-country estimated flows of visitors. Colors of the ribbons correspond to the destination of a trip, while the country of origin is marked with the thin stripe at the end of a ribbon (visualization method based on Krzywinski et al. 2009).

The network partitioning procedure was based on the well-known modularity optimization approach (Newman 2006) using a highly-efficient optimization algorithm recently proposed by Sobolevsky et al. (2013b). In general terms, the procedure assesses the relative strength of particular links versus the estimations of the homogenous null model preserving the strength of each network node. It optimizes the overall modularity score of a network partitioning, which quantifies the strength of intra-cluster connections (in the "ideal" partitioning case they should be as strong as possible) and the weakness of outer ties (supposed to be as weak as possible).

After obtaining the initial split of the network, the partitioning procedure was applied in the iterative manner to the sub-networks inside each community, similarly as in Sobolevsky et al. (2013a). As the result, the Twitter network was split into mobility clusters on three hierarchical levels, each level being a sub-partitioning of the previous one. The initial level (Figure 10A) uncovered four groups of countries that closely agreed with the continental division of the world. In this sense, travel connections e.g. within both Americas were stronger than between America and Europe, while the Europeans traveled more within Europe and Asia than to the other continents. Further sub-divisions followed the same type of common logic, all clusters tend to be spatially connected and well aligned with expected socio-geographical regions. For instance, on level 2 we observed a split into Western, Central, Eastern and Northern Europe (Figure 10B), and on level 3 Central Europe was further divided into the more continental northern part and the Balkans (Table 1). In general, received clusters uncovered that people keep traveling more within their direct 'neighborhoods' rather than connect to further destinations. Furthermore, the fact that clusters were spatially continuous and reflected common regions of the world is in line with the findings of previous studies based on the mobile phone data (Ratti et al. 2010; Blondel et al. 2010; Blondel 2011; Sobolevsky et al. 2013a). It also extends the validity of network-based community detection form a country to global scale. Additionally, we see that the partitioning of mobility networks possess similar regularities compared to human communication networks and might also be used for regional delineation purposes.

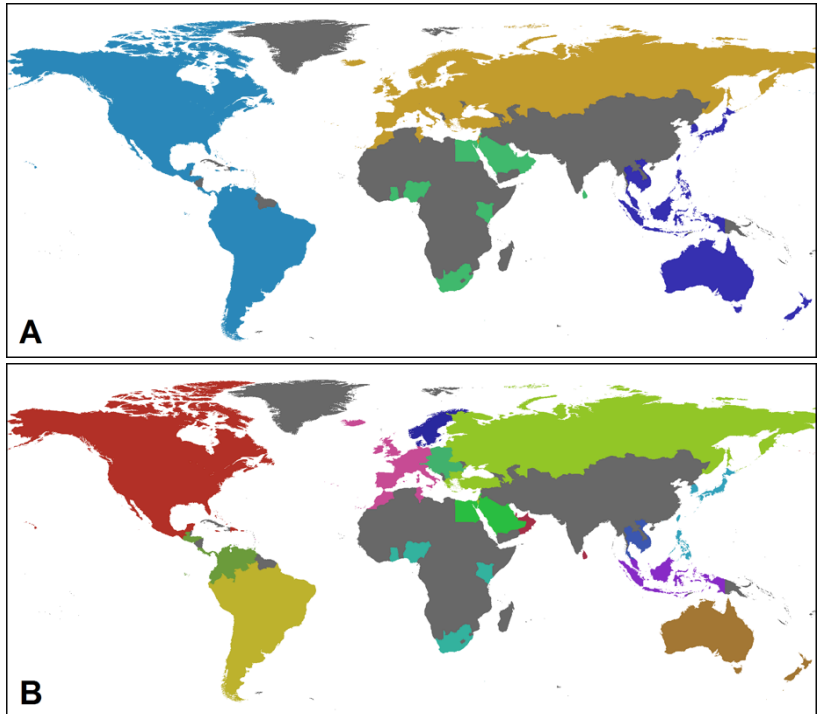

Figure 10. Mobility regions uncovered by the partitioning of country-to-country network of Twitter users flows. Regions distinguished at the first (A) and second (B) level of partitioning. Grey color indicates no data.

Table 1. Countries assigned to different regions of mobility.

| Level 1 | Level 2 | Level 3 | Assigned countries |
|---|---|---|---|
| 1 | 1 | 1 | Bahamas, Canada, Dominican Rep., Jamaica, Mexico, Puerto Rico, United States |
| | 2 | 2 | Colombia, Ecuador, Panama, Trinidad and Tobago, Venezuela |
| | | 3 | Costa Rica, El Salvador, Guatemala, Honduras, |
| | 3 | 4 | Bolivia, Chile, Peru, |
| | | 5 | Argentina, Brazil, Paraguay, Uruguay |
| 2 | 4 | 6 | France, Ireland, Malta, Martinique, Morocco, Portugal, Spain, Tunisia, United Kingdom |
| | | 7 | Belgium, Germany, Iceland, Italy, Luxembourg, Netherlands, Switzerland |
| | 5 | 8 | Denmark, Norway, Sweden |
| | 6 | 9 | Austria, Czech Republic, Hungary, Poland, Romania, Slovakia |
| | | 10 | Bosnia and Herzegovina, Croatia, Kosovo, Macedonia, Serbia, Slovenia |
| | 7 | 11 | Azerbaijan, Bulgaria, Cyprus, Greece, Israel, Kazakhstan, Latvia, Lithuania, Russia, Ukraine |
| | | 12 | Belarus, Estonia, Finland, Turkey |
| 3 | 8 | 13 | Ghana, Nigeria |
| | | 14 | Kenya, South Africa |
| | 9 | 15 | Bahrain, Egypt, Jordan, Kuwait, Lebanon, Saudi Arabia |
| | 10 | 16 | Oman, Qatar, United Arab Emirates |
| | | 17 | Maldives, Sri Lanka |
| 4 | 11 | 18 | Japan, South Korea, Taiwan |
| | | 19 | Philippines |
| | 12 | 20 | Brunei, Indonesia, Malaysia, Singapore |
| | 13 | 21 | Cambodia, Thailand, Vietnam |
| | 14 | 22 | Australia, New Zealand |

**Validation of the results**

In case of global mobility, it is difficult to find the appropriate dataset for the direct validation of results obtained with Twitter, especially one being considerably bias-free in respected to the way it captures human mobility. An interesting comparison could be held for example with the register of flight connections, though possibly hampered by partially disaggregated character of direct connections, often being just a segment of an indirect travel rather than a valid indication of an origin and a destination. In practice such a comparison is also obviously prevented by the difficult accessibility criterion. In this study, we relied on the tourism statistics as provided by the report of the WEF (World Economic Forum 2013) at the country level. We used two of those statistics: international tourist arrivals (thousands, 2011) and international tourism receipts (US$, millions, 2011) and compared them to arrivals estimated with the Twitter data (Figure 11A and 10B). In both cases we found a strong linear correlation (respectively with the $R^2$ of 0.69 and 0.88), which is the confirmation of the validity of received mobility estimations.

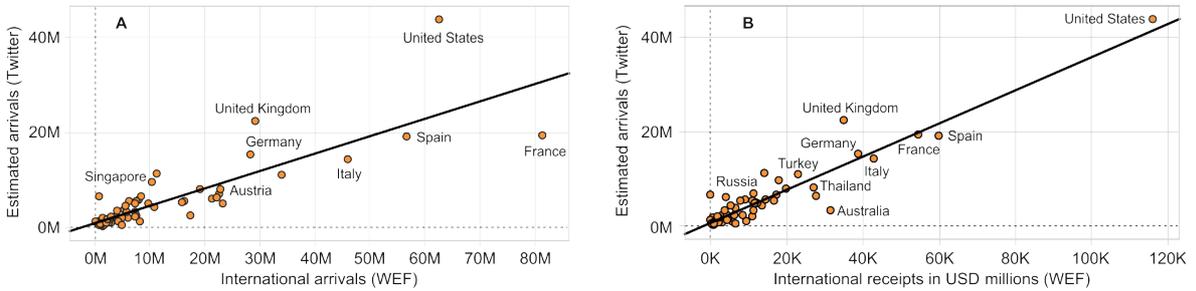

Figure 11. International arrivals estimated with Twitter data versus the arrivals (A) and nominal value of touristic receipts (expenditures by international inbound visitors, B) provided by WEF (2013). Correlation measured with the $R^2$ statistic equals 0.69 and 0.88 respectively.

We further validated the results indirectly by demonstrating that mobility measures derived with Twitter exhibit similar statistical properties as those received on the basis of other datasets. Firstly, we computed the distance between each pair of consecutive user's locations (tweets) and plotted the frequencies of computed displacement on a log-log scale. Similarly to other mobility studies, we found that the distribution is well approximated by the power law (Figure 11A):

$$P(\Delta r) = \Delta r^{-\beta} \quad (2)$$

where Δr is a displacement of certain length and β = 1.62. Importantly, the received exponent stays in the similar range as the results obtained with other mobility datasets such as mobile phone data (González et al. 2008, β = 1.75), bank note dispersal (Brockmann et al. 2006, β = 1.59) and Foursquare check-ins (Cheng et al. 2011, β = 1.88). We also plotted the frequency of previously computed users' radiuses of gyration. As expected, it also followed a power law with an exponent of 1.25.

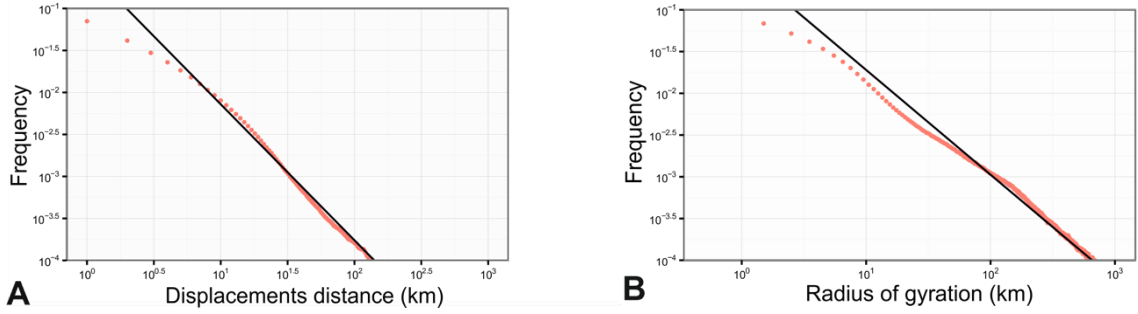
Figure 12. Probability of displacement (A) and frequency of radius of gyration (B).

Considering a limited access to global mobility data suitable for a direct comparison with Twitter-based human flows, we further tested our data against a commonly accepted mobility model, i.e. the classic gravity approach, as yet another way of indirect verification of uncovered patterns. Many studies proved the gravity law to provide a good basis for modelling the intensity of interactions between locations depending on their weights and distance, not only in the context of mobility (e.g. Zipf 1946, Jung et al. 2008, Balcan et al. 2009) but also human interaction networks (Krings et al. 2009, Expert et al. 2011). We assume that if the Twitter data are to be claimed suitable for a description of human mobility, they should follow a similar law and distance dependence as it was proven for other data types e.g. railway connections, airline traffic, mobile telecom records. Furthermore, we tested how much the gravity law holds on a global scale, especially in times when the influence of distance as a friction for mobility is often considered to decrease in importance. We used the gravity model in the form:

$$F_{ij} = A \frac{p_i^{\alpha} p_j^{\beta}}{r_{ij}^{\gamma}} \qquad (3)$$

where $F_{ij}$ represents a flow of people between a pair of countries, $p_i$ is the population in the country of origin, and $p_j$ the population of the destination, $r_{ij}$ is the distance measured between the capitals and A is a constant. To compensate for the limitation of our definition of country-to-country distance we restricted the connections to the countries that are at least 100 km apart. The model was fitted to the two versions of our network. First, the flows were defined as a raw number of Twitter users traveling between two countries, and the population of each country was equal to the number of Twitter residents (Figure 13A). In this case we received the exponents of $\alpha = 0.81$, $\beta = 0.63$ and $\gamma = 1.02$, and an $R^2$ coefficient of 0.79. The second variation of the model (Figure 13B) used the flows estimated based on the Twitter penetration rate (again with the threshold of 0.05%), and the population as provided by the Central Intelligence Agency (2012). Received exponents were fairly similar to the previous results: $\alpha = 0.89$, $\beta = 0.69$ and $\gamma = 1.1$ with only a slight decrease of the $R^2$ coefficient to 0.71.

Received population exponents indicate an underlinear influence of a country's size on the growth of a human flow, both in case of origin and destination, but the influence of the population in a country of origin is obviously bigger. These underlinear relations can be explained with two conjectures. On the one hand, it is plausible that residents of a country do not take part in mobility in an equal manner, instead it is rather a domain of the most active ones. On the other hand, the visitors are never attracted by the whole country but rather only by certain places. On average the number of active users and attractive places among countries may grow slower comparing to their respective total population.

The gamma exponents suggest, as expected, a decrease of interaction intensity with distance (Figure 13A and 13B), however at a slower rate than often pre-assumed, e.g. by Jung et al. (2008) or Krings et al. (2009), $r^2$ decay relationship. This difference can be interpreted through the scale of analysis. In a global scale, where most of the trips are happening by air, increase of a distance comes with relatively smaller effort or cost than in a country or local levels where substantial part of mobility is taking place by land. But even in the world of this subjective 'shrinkage' of distances, certain level of dependency is still preserved. As the reason we suppose social ties, which remain stronger in a local than in a global scale (Takhteyev et al. 2012). In other words, people may simply have more reasons to travel shorter distances, which also correlates well with the results received during network partitioning.

Visually, both variations models seem to be well fitted (Figure 13A and B), with the slopes well reflecting the average tendency in the observed data and the distant decay functions remaining within the standard errors across the whole distance range. The similarity of both models suggests that Twitter data may not only provide a valid picture of mobility of its direct users, but can further be used for the estimation of real human flows.

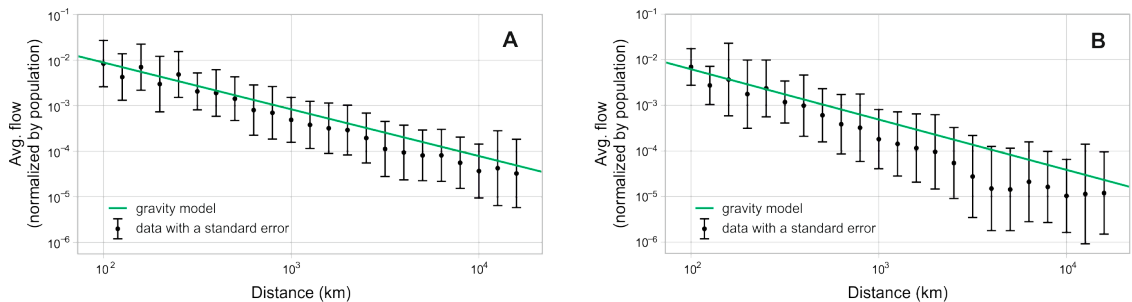

Figure 13. Dependence of human flow ($F_{ij}$) normalized with populations in countries of origin and destination (A $p_i^\alpha p_j^\beta$) on the distance in comparison to a distance decay function ($r^{-\gamma}$) modeled with the gravity law. (A) Network defined based on raw Twitter flows (B) Network of total population flows estimated with the Twitter penetration rate in the country of origin (B).

**Conclusions**

Geo-located Twitter is one of the first free and easily available global data sources that store millions of digital and fully objective records of human activity located in space and time. With our study we demonstrated that, despite the unequal distribution over the different parts of the world and possible bias toward a certain part of the population, in many cases geo-located Twitter can and should be considered as a valuable proxy for human mobility, especially at the level of country-to-country flows. Our approach proposed to capture mobility respective to the nationality of travelers based on the simple yet effective method of assigning each user to her country of residence. In this way, we were able to compare mobility profiles of countries, considering each of them both as a potential origin as well as a destination of international travels. Received results showed that increased mobility measured in terms of probability, diversity of destinations, and geographical spread of travels, is characteristic of more developed countries such as the West-European ones. The travelling distance was additionally affected by the geographic isolation of a country as in the case of Australia or New Zealand. The analysis of temporal patterns indicated the existence of globally universal season of increased mobility at the end of the year, clearly visible regardless of the nationality of travelers. Although the summer mobility was increased for a wide range of countries, it varied in terms of intensity and duration, and in some cases was not visible at all. Additionally, we discovered specific patterns driven either by cultural conditionings or special events occurring in a given country. However in many cases the results just confirmed the common logic expectations, this agreement should be treated as an indicator of the legitimacy of Twitter as a global and objective register of human mobility.

Furthermore, we demonstrated that community detection performed using the Twitter mobility network resulted in spatially cohesive regions that followed the regional division of the world. This finding is important from several stand points. First of all, it is in agreement with the results obtained by other authors (Ratti et al. 2010; Blondel et al. 2010; Sobolevsky et al. 2013a) based on the different types of networks i.e. mobile phone interactions. Furthermore, it extends the spatial validity of the community detection approach from a previously examined country scale to the global one. Finally, it shows that even in the era of globalization and seeming decrease of the influence of distance, people still tend to travel locally, visiting immediate neighboring countries more often than those further away.

As our final contribution we validated, to a certain extent, geo-located Twitter as a proxy of global mobility behavior. First of all, we demonstrated that volumes of visitors estimated for different countries based on Twitter data are in line with the official statistics on international tourism. The correlation ($R^2$ around 0.7) shows a fairly good correspondence given the wider scope of mobility captured through Twitter, as well as significantly different character of data acquisition. Secondly, we confirmed that Twitter data exhibits similar statistical properties as other mobility datasets. For instance, the measures such as a radius of gyration and the probability of displacement are well approximated with the power law distribution, similarly as in Brockmann et al. (2006) or Cheng et al. (2011), and the network of the estimated flows of travelers can be well approximated with the classic model of a mobility i.e. the gravity model. Altogether, we believe that our analysis proved the potential of geo-located Twitter as a fully objective, and freely accessible source for global mobility studies. Further research will focus on the exploration how far this potential can be translated to finer spatial scales.

**Acknowledgements**

*This research was funded by the Austrian Science Fund (FWF) through the Doctoral College GIScience (DK W 1237-N23), Department of Geoinformatics - Z_GIS, University of Salzburg, Austria. We would like to thank Sebastian Grauwin and Karolina Stanislawska for their support. We further thank the MIT SMART Program, the Center for Complex Engineering Systems (CCES) at KACST and MIT CCES program, the National Science Foundation, the MIT Portugal Program, the AT&T Foundation, Audi Volkswagen, BBVA, The Coca Cola Company, Ericsson, Expo 2015, Ferrovial, GE and all the members of the MIT Senseable City Lab Consortium for supporting the research.*